\documentclass[aps,prb,twocolumn,floatfix,superscriptaddress]{revtex4-1}
\usepackage{graphicx,txfonts,bm}
\usepackage[colorlinks,citecolor=magenta,linkcolor=blue]{hyperref}

\begin{document}


\title{Protracted Kondo screening and kagome bands in heavy-fermion metal Ce$_{3}$Al}


\author{Li Huang}
\email{lihuang.dmft@gmail.com}
\affiliation{Science and Technology on Surface Physics and Chemistry Laboratory, P.O. Box 9-35, Jiangyou 621908, China}

\author{Haiyan Lu}
\affiliation{Science and Technology on Surface Physics and Chemistry Laboratory, P.O. Box 9-35, Jiangyou 621908, China}

\date{\today}


\begin{abstract}
Ce$_{3}$Al is an archetypal heavy-fermion compound with multiple crystalline phases. Here, we try to investigate its electronic structures in the hexagonal phase ($\alpha$-Ce$_{3}$Al) and cubic phase ($\beta$-Ce$_{3}$Al) by means of a combination of density functional theory and single-site dynamical mean-field theory. We confirm that the 4$f$ valence electrons in both phases are itinerant, accompanied with strong valence state fluctuations. Their 4$f$ band structures are heavily renormalized by electronic correlations, resulting in large effective electron masses. The Kondo screening in Ce$_{3}$Al would be protracted over a wide range of temperature since the single-impurity Kondo temperature $T_{K}$ is much higher than the coherent Kondo temperature $T^{*}_{K}$. Especially, the crystal structure of $\alpha$-Ce$_{3}$Al forms a layered kagome lattice. We observe conspicuous kagome-derived flat bands and Dirac cones (or gaps) in its quasiparticle band structure. Therefore, it is concluded that the hexagonal phase of Ce$_{3}$Al will be a promising candidate of heavy-fermion kagome metal.         
\end{abstract}


\maketitle

\section{introduction\label{sec:intro}}

The discovery of fully gapped $d$-wave superconductivity in CeCu$_{2}$Si$_{2}$ has triggered a great deal of interests in heavy-fermion materials~\cite{PhysRevLett.43.1892,Stockert2011,Pang5343,Yamashitae1601667,Yuan2104}. These materials are typically intermetallic compounds containing rare earths (Ce, Sm, and Yb) or actinides (U, Np, and Pu). They exhibit a barrage of unusual properties that we still don't fully understand. For example, in comparison to normal metals, they have enormous values of effective electron masses $m^{*}$, linear specific heat coefficients $\gamma$, and low-temperature magnetic susceptibilities $\chi$~\cite{RevModPhys.56.755}. Even more fascinating, the heavy-fermion materials host a plethora of atypical phenomena and exotic quantum states, including quantum criticality, unconventional superconductivity, non-Fermi-liquid state, topological Kondo insulator and topological Kondo semimetal~\cite{Lai93,dzero014749,RevModPhys.73.797,PhysRevLett.104.106408,Schuberth485,Gegenwart2008}, just to name a few. Due to these tantalizing properties, the search for and characterization of heavy-fermion materials have become a rapid growing field in the condensed matter physics. 

It is worth noting that quite a large portion of heavy-fermion materials are cerium-based intermetallic compounds~\cite{Weng_2016}. They have attracted more attentions than the other uranium-based and ytterbium-based heavy-fermion compounds~\cite{RevModPhys.56.755}. For example, CeCu$_{2}$Si$_{2}$ is the first known heavy-fermion superconductor~\cite{PhysRevLett.43.1892}, as mentioned before. However, after 40 years, the exact symmetry of its superconducting gap has not been settled yet~\cite{PhysRevLett.121.157004,PhysRevLett.120.217001,PhysRevLett.112.067002}. CeCoIn$_{5}$ is another heavy-fermion superconductor with a relatively high $T_{c}$ ($T_c > 2.3$~K). Besides unconventional superconductivity, it also exhibits spin-density-wave magnetic ordering~\cite{PhysRevX.6.041059}, Fulde-Ferrell-Larkin-Ovchinnikov state~\cite{PhysRevLett.96.117001}, and antiferromagnetic quantum critical point~\cite{PhysRevLett.97.106606,PhysRevB.99.054506,PhysRevLett.117.016601}, which can be easily modulated by external conditions, such as temperature, pressure, magnetic field, and chemical doping. Recently, it is used as a testbed to unveil the quasiparticle dynamics (collective hybridization between localized 4$f$ moments and conduction electrons) at $T > T^{*}$, where $T^{*}$ marks the heavy-fermion coherence temperature~\cite{PhysRevLett.124.057404}. The third example concerns the cerium-based heavy-fermion topological semimetals, which provide a promising setting to study topological semimetals~\cite{RevModPhys.90.015001} driven by electronic correlations. For instance, Ce$_{3}$Bi$_{4}$Pd$_{3}$, a heavy-fermion system without centrosymmetry and magnetic order, is newly found to be a realization of the so-called Weyl-Kondo semimetal~\cite{Lai93,PhysRevLett.118.246601,PhysRevB.101.075138}, which features strongly renormalized Weyl nodes in the bulk and hosts Fermi arcs on the surface.  

\begin{figure}[th]
\centering
\includegraphics[width=\columnwidth]{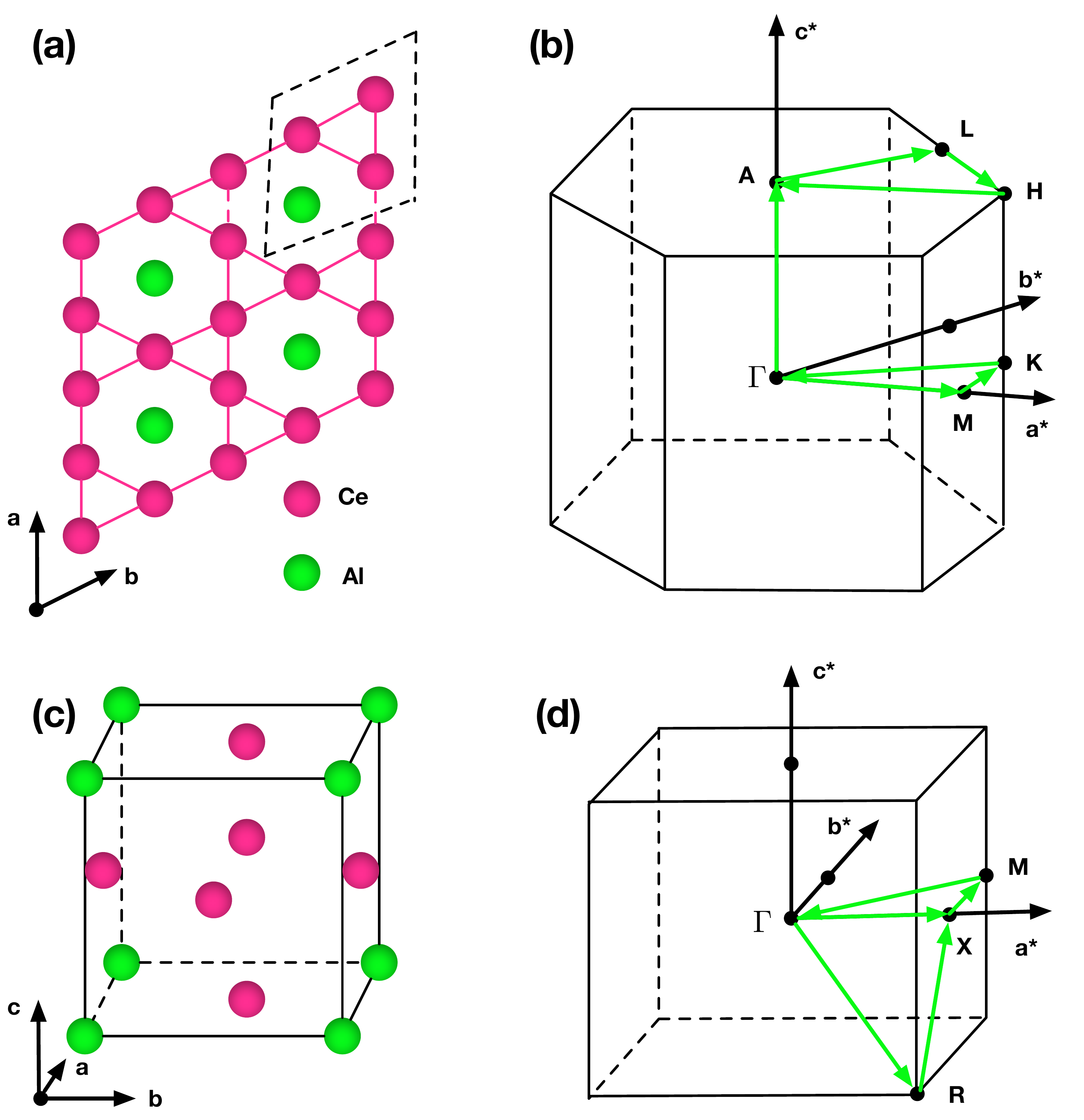}
\caption{(Color online). (a) and (c) Crystal structures of $\alpha$- and $\beta$-Ce$_{3}$Al. Here, Ce and Al atoms are represented by red and green balls, respectively. In panel (a), only the kagome layer is illustrated, and the dashed rhomboid means the unit cell. (b) and (d) Schematic pictures for Brillouin zones of $\alpha$- and $\beta$-Ce$_{3}$Al. Some selected high-symmetry directions are labelled. \label{fig:bz}}
\end{figure}

Now let us turn to the cerium-aluminium heavy-fermion system Ce$_{x}$Al$_{y}$. It contains at least four stable intermetallic compounds, specifically, CeAl$_{y}$ ($y = 1$, 2, 3) and Ce$_{3}$Al. The ground states of CeAl and CeAl$_{2}$ are antiferromagnetic, which are quite complicated. Below 3.5~K, CeAl$_{2}$ enters an antiferromagnetic phase, in which the magnitudes and directions of the magnetic moments exhibit spatial periodicity. It is generally believed that this spin-density-wave-like antiferromagnetic ordering develops out of hybridization between localized 4$f$ spins and conduction electrons (Kondo effect)~\cite{Schweizer_2008}. CeAl$_{3}$ is known to be the first discovered heavy-fermion metal with tremendous magnitude of the linear specific heat term [$\gamma = 1620$~mJ / (mole $\cdot$ K$^2$)]~\cite{PhysRevLett.35.1779}. Its ground state is also antiferromagnetic with $T_N$ =1.2~K~\cite{nakamura2644}. As for Ce$_{3}$Al, at room temperature it crystallizes into a hexagonal Ni$_{3}$Sn-type structure [i.e. $\alpha$-Ce$_{3}$Al, see Fig.~\ref{fig:bz}(a)]. Above 500~K, it transforms into a cubic Cu$_{3}$Au-type structure [i.e. $\beta$-Ce$_{3}$Al, see Fig.~\ref{fig:bz}(c)]. Below 115~K, another structural phase transition occurs and the monoclinic $\gamma$ phase appears. The antiferromagnetic ordering develops below 2.5~K~\cite{SERA198782}. Previous studies have revealed fingerprints of heavy fermions in the thermodynamic, magnetic, and transport properties in Ce$_{3}$Al~\cite{PhysRevB.40.10766,PhysRevB.55.5937,li367575}. Strictly, Ce$_{3}$Al is classified as a heavy-fermion system because of the large linear specific heat coefficient [$\gamma = 85 \sim 114$~mJ / (mole Ce $\cdot$ K$^2$)]~\cite{Singh_2014}, but its 4$f$ electrons are not as heavy as those in CeAl$_{y}$~\cite{PhysRevB.55.5937,PhysRevB.40.10766}. 

Among the intermetallic compounds made by Ce and Al, CeAl$_{2}$ and CeAl$_{3}$ have been most extensively studied in relation to their magnetic and heavy-fermion properties~\cite{Schweizer_2008,PhysRevLett.35.1779,nakamura2644}. On the other hand, only a few works have been reported for Ce$_{3}$Al~\cite{SERA198782,PhysRevB.40.10766,PhysRevB.55.5937,li367575}. Consequently, we know a little about the detailed electronic structures of Ce$_{3}$Al. The variation of the 4$f$ electronic states across the three phases remains unknown so far. Therefore, in the present work, we make a systematic study about the electronic structures of the high-symmetry phases of Ce$_{3}$Al ($\alpha$-Ce$_{3}$Al and $\beta$-Ce$_{3}$Al) by means of a state-of-the-art first-principles many-body approach. We try to calculate the quasiparticle band structures, density of states, self-energy functions, and valence state histograms of both phases. The Kondo temperatures and effective electron masses are also measured. Our results discover that the 4$f$ electrons in both phases are itinerant and fluctuating heavily among various electronic configurations, but they are not coherently screened by conduction electrons to form a Fermi-liquid state~\cite{Nozieres1998} in the temperatures that we are interested in. In addition, it is suggested that $\alpha$-Ce$_{3}$Al is a prototype of kagome metal, which is characterized by the coexistence of dispersionless and linearly dispersive energy bands~\cite{PhysRevB.45.12377}.  

The rest of this paper is organized as follows. Firstly, the computational details are introduced in section~\ref{sec:method}. Then in section~\ref{sec:results}, the major results, including the momentum-resolved spectral functions, density of states, Kondo temperatures, self-energy functions, and valence state histograms of the two phases are presented and analyzed. Section~\ref{sec:dis} is devoted to discuss the kagome-derived flat bands and Dirac cones (or gaps) in the quasiparticle band structure of $\alpha$-Ce$_{3}$Al. Finally, section~\ref{sec:summary} serves as a short summary. 

\section{method\label{sec:method}}

In the present work, we employed the single-site dynamical mean-field theory in combination with the density functional theory (dubbed DFT + DMFT)~\cite{RevModPhys.78.865,RevModPhys.68.13} to study the electronic structures of $\alpha$- and $\beta$-Ce$_{3}$Al. Notice that this method has been successfully applied to explore the physical properties of many cerium-based heavy-fermion materials~\cite{Goremychkin186,Shim1615,PhysRevB.99.045122,PhysRevB.98.195102,PhysRevB.81.195107,PhysRevB.94.075132}. 

The DFT calculations were done by using the \texttt{WIEN2K} code, which implements a full-potential linearized augmented plane-wave formalism~\cite{wien2k}. We adopted the experimental crystal structures~\cite{PhysRevB.40.10766,li367575}. The muffin-tin radii for Ce and Al atoms were 2.5 and 2.4 a.u, respectively. $R_{\text{MT}} \times K_{\text{MAX}} = 7.0$. The Perdew-Burke-Ernzerhof functional~\cite{PhysRevLett.77.3865}, i.e. the generalized gradient approximation was adopted to express the exchange-correlation potential. The $k$-meshes for Brillouin zone integrations for the $\alpha$- and $\beta$-Ce$_{3}$Al phases were 15 $\times $ 15 $\times$ 16 and 15 $ \times $ 15 $\times $ 15, respectively. The spin-orbit coupling was considered in all calculations explicitly. 

Since the electronic correlations among Ce's 4$f$ valence electrons are crucial, we have to taken them into accounts. We utilized the DMFT method to treat the correlated nature of 4$f$ electrons. Here, we used the \texttt{eDMFT} code developed by K. Haule~\cite{PhysRevB.81.195107}. The system temperatures for $\alpha$- and $\beta$-Ce$_{3}$Al were set to be $T \approx 230$~K ($\beta = 50.0$) and $T \approx 580$~K ($\beta = 20.0$), respectively. A large energy window (from -10~eV to +10~eV with respect to the Fermi level) was used to construct the DMFT projectors and local orbitals. The Coulomb repulsion interaction parameter $U$ and Hund's exchange interaction parameter $J_{\text{H}}$ for Ce's 4$f$ electrons were 6.0~eV and 0.7~eV, respectively~\cite{Shim1615,PhysRevB.99.045122,PhysRevB.98.195102,PhysRevB.81.195107,PhysRevB.94.075132}. The double counting term for the self-energy functions was subtracted via the exact scheme~\cite{PhysRevLett.115.196403}. In order to solve the auxiliary quantum impurity problems for 4$f$ electrons, a hybridization expansion continuous-time quantum Monte Carlo impurity solver (dubbed CT-HYB)~\cite{PhysRevB.75.155113,PhysRevLett.97.076405,RevModPhys.83.349} was used. For each quantum impurity solver run, the number of Monte Carlo steps was up to 200 million per CPU process. We performed charge fully self-consistent DFT + DMFT calculations. In order to obtain good convergence, the maximum number of DFT + DMFT iterations was set to 100. 

\section{results\label{sec:results}}

\begin{table*}[ht]
\caption{Important model parameters for the electronic structures of $\alpha$-Ce$_{3}$Al and $\beta$-Ce$_{3}$Al. They include the width of conduction band below the Fermi level ($W$), the averaged 4$f$ electron level ($\epsilon_f$), the imaginary part of hybridization function at the Fermi level [$\text{Im}\Delta(E_F)$], the density of states of conduction electrons at the Fermi level [$\rho_c(E_F)$], the Kondo temperature ($T_K$), the coherent Kondo temperature ($T^{*}_{K}$), the quasiparticle weights ($Z_{5/2}$ and $Z_{7/2}$), the effective electron masses ($m^{*}_{5/2}$ and $m^{*}_{7/2}$), and probabilities of $4f^{0}$, $4f^{1}$, and $4f^{2}$ configurations ($p_0$, $p_1$, and $p_2$). See main texts for more details. \label{tab:ratio}}
\begin{ruledtabular}
\begin{tabular}{rccccccccccccc}
cases & $W$ & $\epsilon_f$\footnotemark[1] & $\text{Im}\Delta(E_F)$\footnotemark[2] & $\rho_c(E_F)$  & $T_{K}$ & $T^{*}_{K}$ & $Z_{5/2}$ & $Z_{7/2}$ & $m^{*}_{5/2}$ & $m^{*}_{7/2}$ & $p_0$ & $p_1$ & $p_2$ \\
\hline
$\alpha$-Ce$_{3}$Al & 7.05~eV & -1.60~eV  & -0.147~eV & 8.37~eV$^{-1}$  & 684~K & 56~K & 0.17 & 0.41 & 5.88$m_e$ & 2.45$m_e$ & 8.56\% & 85.36\% & 6.01\%\\
$\beta$-Ce$_{3}$Al  & 6.60~eV & -1.46~eV  & -0.149~eV & 4.23~eV$^{-1}$  & 885~K & 47~K & 0.13 & 0.27 & 7.75$m_e$ & 2.56$m_e$ & 7.21\% & 87.78\% & 4.96\%\\
\end{tabular}
\end{ruledtabular}
\footnotetext[1]{$\epsilon_f = [\epsilon_{f}(j = 5/2) + \epsilon_{f}(j=7/2)]/2$.}
\footnotetext[2]{Only for the $4f_{5/2}$ states.}
\end{table*}

\begin{figure*}[ht]
\centering
\includegraphics[width=\textwidth]{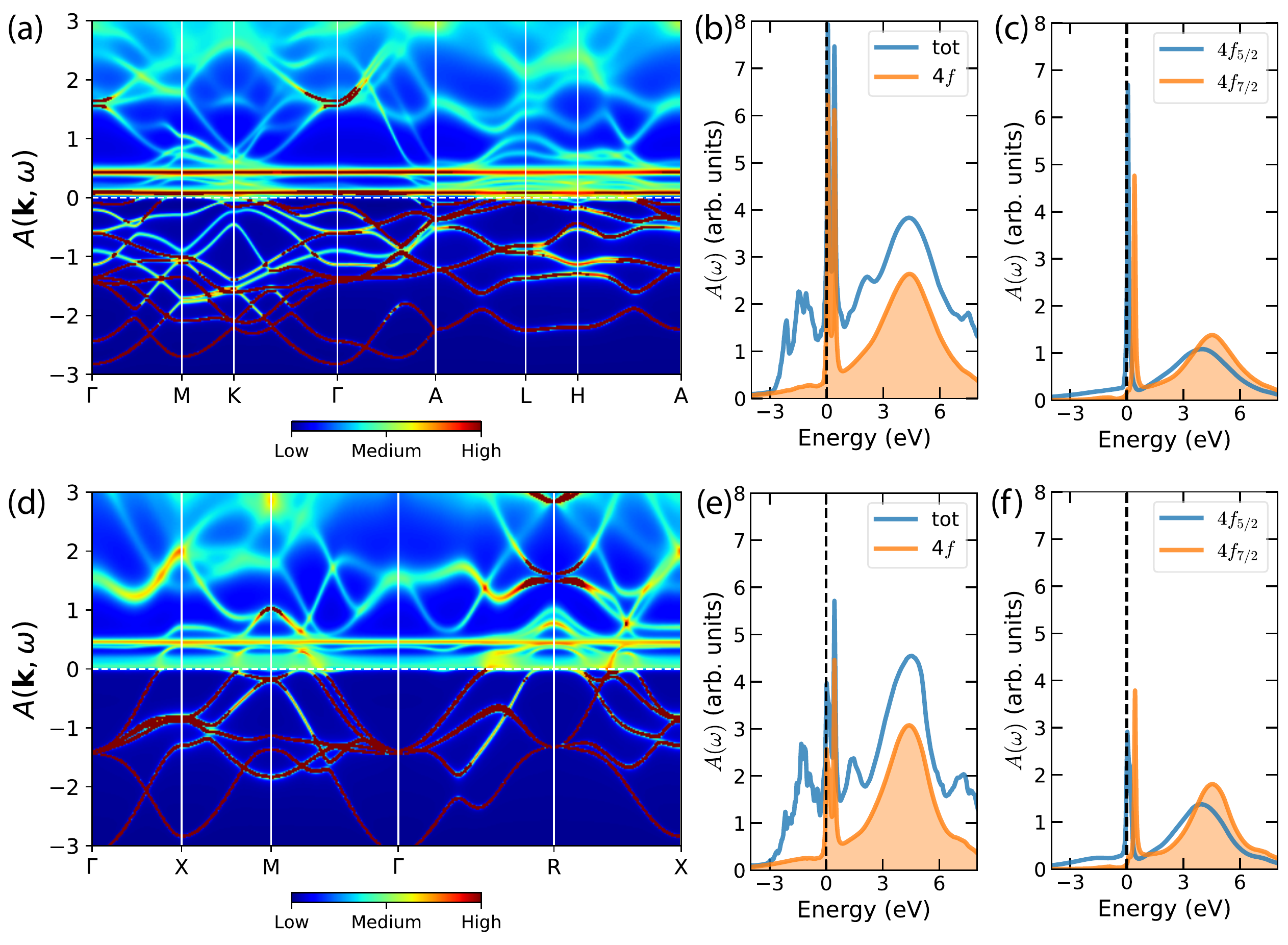}
\caption{(Color online). Quasiparticle band structures of Ce$_{3}$Al obtained via DFT + DMFT calculations. The results for the $\alpha$ and $\beta$ phases of Ce$_{3}$Al are plotted in the upper and lower panels, respectively. (a) and (d) Momentum-resolved spectral functions $A(\mathbf{k},\omega)$. (b) and (e) Total and 4$f$ partial density of states [i.e. $A(\omega)$ and $A_{4f}(\omega)$]. (c) and (f) $j$-resolved 4$f$ partial density of states [i.e. $A_{4f_{5/2}}(\omega)$ and $A_{4f_{7/2}}(\omega)$]. Here, the horizontal or vertical dashed lines denote the Fermi level. The data of $A(\omega)$, $A_{4f}(\omega)$, $A_{4f_{5/2}}(\omega)$, and $A_{4f_{7/2}}(\omega)$ are rescaled for a better view. \label{fig:akw}}
\end{figure*}

\begin{figure*}[ht]
\centering
\includegraphics[width=\textwidth]{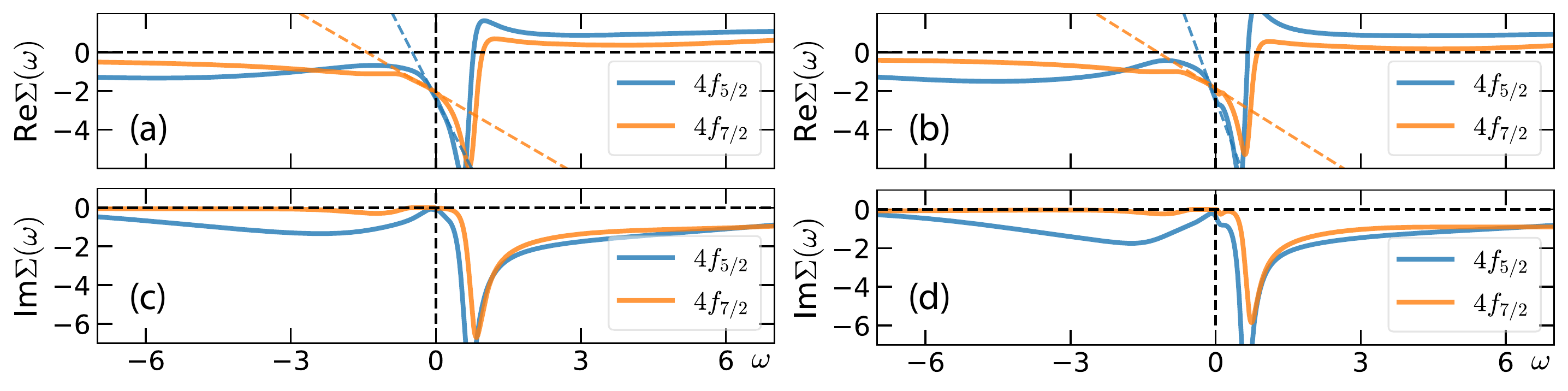}
\caption{(Color online). Real and imaginary parts of self-energy functions of Ce-4$f$ electrons at real axis. (a) and (c) $\alpha$-Ce$_{3}$Al. (b) and (d) $\beta$-Ce$_{3}$Al. In panels (a) and (b), the blue and yellow dashed lines are linear fitting for the low-frequency quasi-linear regimes of the self-energy functions. See main text for more explanations. \label{fig:sig}}
\end{figure*}

\begin{figure}[ht]
\centering
\includegraphics[width=\columnwidth]{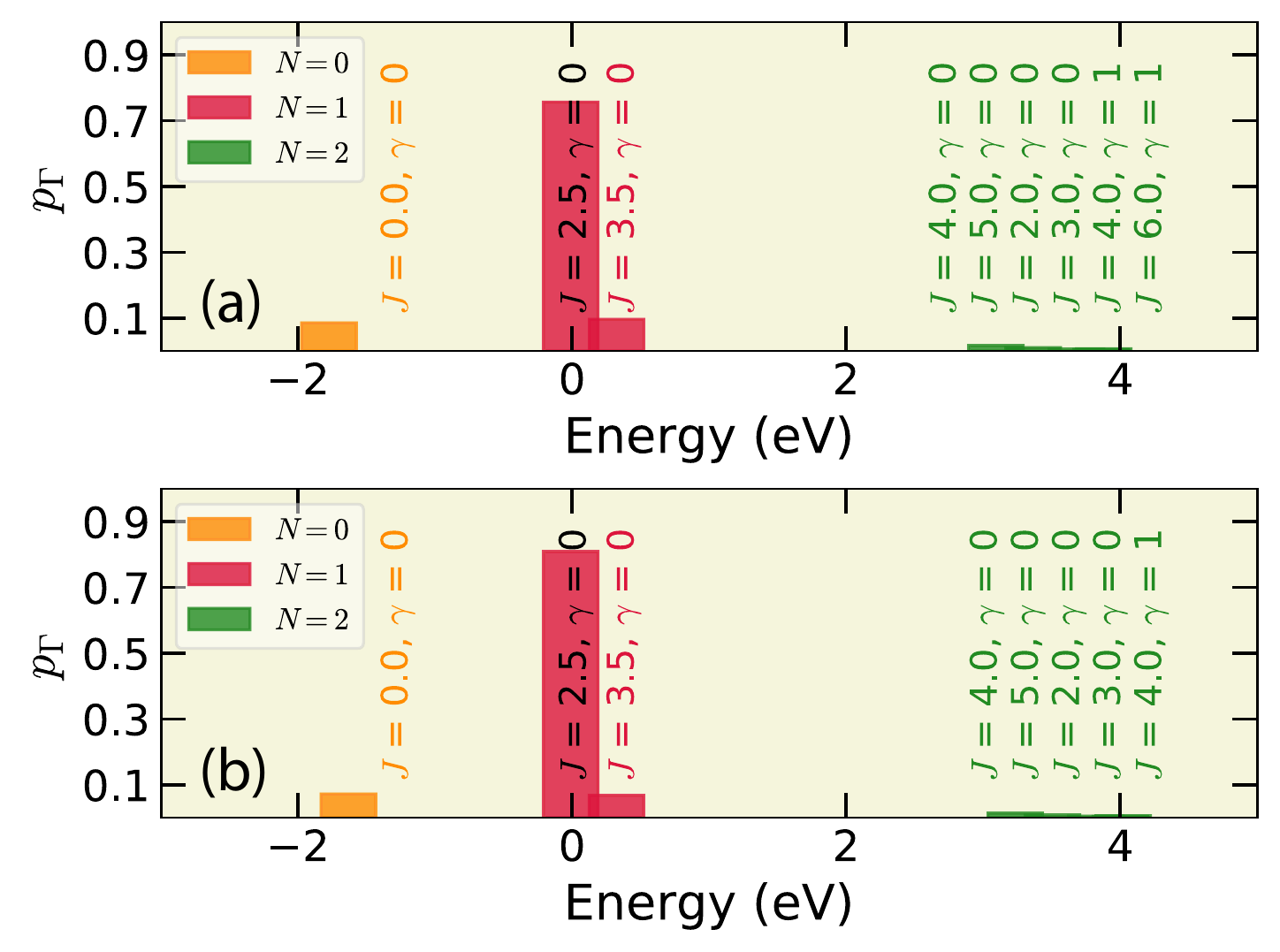}
\caption{(Color online). Valence state histograms of Ce's 4$f$ electrons. The data for $4f^{3}$ ($N = 3$) configurations are too trivial to be seen in this figure. (a) $\alpha$-Ce$_{3}$Al. (b) $\beta$-Ce$_{3}$Al. \label{fig:prob}}
\end{figure}

\emph{Quasiparticle band structures.} The momentum-resolved spectral functions (or equivalently quasiparticle band structures) $A(\mathbf{k},\omega)$ along some selected high-symmetry directions for the $\alpha$ and $\beta$ phases of Ce$_{3}$Al are visualized in Fig.~\ref{fig:akw}(a) and (d), respectively. The spectra of the two phases share some common characteristics. At first, the spectra below the Fermi level look quite coherent and have large dispersions, which are attributed to the contributions of conduction electrons. Secondly, the incoherent 4$f$ bands dominate above the Fermi level, resulting in blur and diffused spectra. Thirdly, we observe remarkable stripe-like features near the Fermi level. One stripe pins exactly at the Fermi level, while the other is a few hundred meV above the Fermi level. These stripes are related with the spin-orbit splitting $4f_{5/2}$ and $4f_{7/2}$ bands, which have been seen in Ce metal~\cite{PhysRevB.99.045122} and some other cerium-based heavy-fermion compounds, such as Ce$T$In$_{5}$~\cite{PhysRevB.81.195107,Shim1615}, CeIn$_{3}$~\cite{PhysRevB.94.075132}, CeB$_{6}$~\cite{PhysRevB.95.155140}, and Ce$M$$_{2}$Si$_{2}$~\cite{PhysRevB.98.195102}. The low-lying $4f_{5/2}$ bands are extremely flat and intense, leading to sharp peaks at the Fermi level in the density of states. They signal the itinerant behaviors of $4f$ electrons in $\alpha$- and $\beta$-Ce$_{3}$Al.

Fig.~\ref{fig:akw}(b) and (e) depict the total density of states $A(\omega)$ and 4$f$ partial density of states $A_{4f}(\omega)$. $A_{4f}(\omega)$ shows prominent three-peaks structure, which is typical in strongly correlated metals~\cite{RevModPhys.68.13}. The Absikosov-Suhl-like quasiparticle peak (or Kondo resonance peak) is well developed at the Fermi level. Due to the spin-orbit coupling effect, it is split into two sub-peaks, which are counterparts of the stripe-like structures as seen in $A(\mathbf{k},\omega)$. In addition, there is a broad and smooth ``hump'' at 2~eV $< \omega <$ 8~eV. It is mainly assigned to the upper Hubbard bands of cerium's 4$f$ orbitals. On the other hand, the lower Hubbard bands (residing from -3~eV to -0.5~eV) exhibit small spectral weights, which is consistent with the fact that most of cerium's 4$f$ orbitals are unoccupied. In Fig.~\ref{fig:akw}(c) and (f), we further show the $j$-resolved 4$f$ partial density of states. We find that the energy separation between the $4f_{5/2}$ and $4f_{7/2}$ peaks is about 350~meV, which is somewhat larger than those in the other cerium-based heavy-fermion compounds~\cite{PhysRevB.81.195107,PhysRevB.94.075132,PhysRevB.98.195102}. Here we are going to introduce a new variable $R$. It is the ratio of heights of the two spin-orbit splitting peaks, i.e. $R \equiv h(4f_{5/2}) / h(4f_{7/2})$. We find that $R > 1.0$ for $\alpha$-Ce$_{3}$Al, while $R < 1.0$ for $\beta$-Ce$_{3}$Al. This difference can be easily explained by the temperature effect. When the temperature is increased, the 4$f$ electrons should become more and more incoherent~\cite{Goremychkin186,Shim1615}. Thus, the low-lying $4f_{5/2}$ bands are suppressed, so does $R$. 

\emph{Kondo screening.} The physical properties of heavy-fermion materials are controlled by two essential factors~\cite{RevModPhys.56.755}. One is the electronic correlations among the localized $f$ electrons, another one is the hybridization (coupling) between localized $f$ electrons (localized $f$ moments) and conduction electrons. In the weak hybridization limit, the Ruderman-Kittel-Kasuya-Yosida (RKKY) interaction dominates, which drives the localized $f$ electrons to yield magnetic ordering states. On the other hand, when the hybridization is strong, the Kondo mechanism begins to work. The localized moments of $f$ electrons are screened by conduction electrons, and a heavy Fermi-liquid state appears at low temperature. Traditionally, the Kondo temperature is defined to mark the energy scale of screening of a localized $f$ electron. We can use the following equation to make a rough estimation about the single-impurity Kondo temperature $T_{K}$~\cite{PhysRevB.64.155111}:
\begin{equation}
\label{eq:tk}
T_K = \sqrt{W |\text{Im} \Delta(E_F)|}\exp{\left(-\frac{\pi |\epsilon_f|}{2N_f |\text{Im} \Delta(E_F)|}\right)}.
\end{equation}
Here $W$ means that width of occupied conduction bands, $\text{Im}\Delta(E_F)$ means the imaginary part of hybridization function at the Fermi level, $\epsilon_f$ is the averaged impurity level for $4f_{5/2}$ and $4f_{7/2}$ states, $N_{f}$ is the band degeneracy of $f$ electrons. Since the Kondo resonance peak in the Fermi level is mainly associated with the six-fold $4f_{5/2}$ states, $N_f = 6$ in Eq.~(\ref{eq:tk}). We can further evaluate the coherent Kondo temperature (or protracted Kondo screening temperature) $T^{*}_{K}$, which indicates the temperature that all localized $f$ electrons are screened by conduction electrons. Note that $T^{*}_K$ is related to $T_{K}$ via the following formula~\cite{Nozieres1998,pz:2019}:
\begin{equation}
\label{eq:tk1}
T^{*}_{K} = \frac{\rho_c(E_F)}{N_f} T^2_{K},
\end{equation}
where $\rho_c(E_F)$ denotes the density of states of conduction electrons at the Fermi level. We tried to calculate $T_K$ and $T^{*}_{K}$ for $\alpha$- and $\beta$-Ce$_{3}$Al with Eq.~(\ref{eq:tk}) and (\ref{eq:tk1}). The calculated results, together with the necessary parameters, are summarized in Table~\ref{tab:ratio}. Nozi\'{e}res' exhaustion theory~\cite{Nozieres1998} argues that only those conduction electrons within the Kondo energy scale around the Fermi level $E_F$ can contribute to the screening. As a consequence, the coherent Kondo temperature $T^{*}_{K}$ could be much lower than the single-impurity Kondo temperature $T_K$ in heavy-fermion materials with low carrier density~\cite{Luo13520}. Our results clearly reveal that $T_{K} \gg T^{*}_{K}$, which suggests that both phases are in the protracted Kondo screening states ($T_{K} > T > T^{*}_{K}$)~\cite{pz:2019}. The low-temperature resistivity of $\alpha$-Ce$_{3}$Al exhibits two characteristic maximums around 100~K and 3~K, and a minimum around 20~K. The high-temperature anomaly may be due to the $\alpha-\gamma$ phase transition, and the low-temperature anomaly may originate from the occurrence of antiferromagnetic order. As for the resistance minimum, previous studies suggested that it marks the onset of the Kondo screening (Kondo minimum)~\cite{SERA198782,PhysRevB.40.10766,Singh_2014}. Instead, we believe that it denotes a phase transition from the protracted Kondo screening state to the coherent Kondo screening state.     

%
%
%
%

\emph{Self-energy functions.} Next, let us examine the heavy-fermion states in Ce$_{3}$Al. At first, we have to perform analytical continuation for the self-energy functions at Matsubara axis $\Sigma(i\omega_n)$ via the maximum entropy method~\cite{jarrell}. The obtained self-energy functions at real axis $\Sigma(\omega)$ are shown in Fig.~\ref{fig:sig}. We find that the orbital differentiation in the self-energy functions is considerable. For the imaginary parts of self-energy functions $\text{Im}\Sigma(\omega)$, sizable gaps occur at $\omega = 0$ for the $4f_{7/2}$ states, indicating their insulating-like nature. However, $|\text{Im}\Sigma(\omega)|$ for the $4f_{5/2}$ states is finite. It means the existence of low-energy electron scattering. For the real parts of self-energy functions $\text{Re}\Sigma(\omega)$, we observe quasi-linear behaviors in the vicinity of the Fermi level. Therefore, we can utilize the following equation to evaluate the quasiparticle weights $Z$ and effective electron masses $m^{*}$~\cite{RevModPhys.68.13}:
\begin{equation}
Z^{-1} = \frac{m^{*}}{m_e} = 1 - \frac{\partial \text{Re}\Sigma(\omega)}{\partial \omega}\Big|_{\omega = 0}.
\end{equation}
The calculated results are tabulated in Table~\ref{tab:ratio}. Just as expected, the 4$f$ electron correlation is orbital-dependent. The low-lying $4f_{5/2}$ states are more correlated than the high-lying $4f_{7/2}$ states ($Z_{5/2} < Z_{7/2}$), so the 4$f$ electrons in the $j = 5/2$ states suffer more renormalization and become heavier. It is not surprised that the 4$f$ electrons in Ce$_{3}$Al are much lighter than those in the other Ce-Al intermetallic heavy-fermion compounds. After all $\gamma_{\text{Ce}_{3}\text{Al}} / \gamma_{\text{CeAl}_{3}} \approx 0.2$~\cite{PhysRevLett.35.1779,Singh_2014}. 

\emph{Valence state histograms.} Valence fluctuation and mixed-valence behavior are general features in correlated $f$-electron systems~\cite{PhysRevB.81.195107,PhysRevB.99.045122}. Especially, when the $f$ electrons reside in the itinerant side (strong hybridization limit), these effects will become more prominent~\cite{PhysRevB.99.045122}. In this work, we tried to calculate the valence state histograms (i.e. atomic eigenstates probabilities) of Ce$_{3}$Al to quantify its valence fluctuation~\cite{PhysRevB.75.155113}. See Fig.~\ref{fig:prob} for the histograms. Here, we use three good quantum numbers to label these atomic eigenstates, i.e., total occupancy $N$, total angular momentum $J$, and $\gamma$ which stands for the rest of the atomic quantum numbers, such as $J_z$. We have the following findings. At first glance, the distributions of valence state histograms of $\alpha$- and $\beta$-Ce$_{3}$Al are quite similar. They are diverse. Accordingly, the valence state fluctuations in both phases are intense. Second, the probability of $4f^{1}$ configuration is undoubtedly overwhelming. But the contributions from the other configurations, such as $4f^{0}$ and $4f^{2}$, are also crucial (see Table~\ref{tab:ratio}). The contributions from $4f^{3}$ and those with more 4$f$ occupancies are trivial. With these information, we can easily calculate the expected 4$f$ occupancy $\langle N \rangle$ and expected total angular momentum $\langle J \rangle$. For $\alpha$-Ce$_{3}$Al, $\langle N \rangle \approx 0.976$ and $\langle J \rangle \approx 2.47$. While for $\beta$-Ce$_{3}$Al, $\langle N \rangle \approx 0.978$ and $\langle J \rangle \approx 2.46$. These data are quite close. Third, the ground states of both phases are $|N = 1, J = 2.5, \gamma = 0 \rangle$, which is the same with Ce and most of the cerium-based heavy-fermion systems~\cite{PhysRevB.94.075132,PhysRevB.95.155140,PhysRevB.98.195102}. It accounts for 75.7\% for $\alpha$-Ce$_{3}$Al and 80.9\% for $\beta$-Ce$_{3}$Al. Finally, note that the data presented here are quite analogous to those of $\alpha$-Ce ($\langle N \rangle \approx 0.98$, $\langle J \rangle \approx 2.45$, and the probability of $|N = 1, J = 2.5, \gamma = 0 \rangle$ is approximately 68.31\%)~\cite{PhysRevB.99.045122,PhysRevB.75.155113}, in which the 4$f$ electrons are known to be itinerant and strongly hybridize with conduction electrons. So, to some extent, the 4$f$ electronic structures of Ce$_{3}$Al and $\alpha$-Ce are similar. 

\section{discussion\label{sec:dis}}

\begin{figure*}[ht]
\centering
\includegraphics[width=\textwidth]{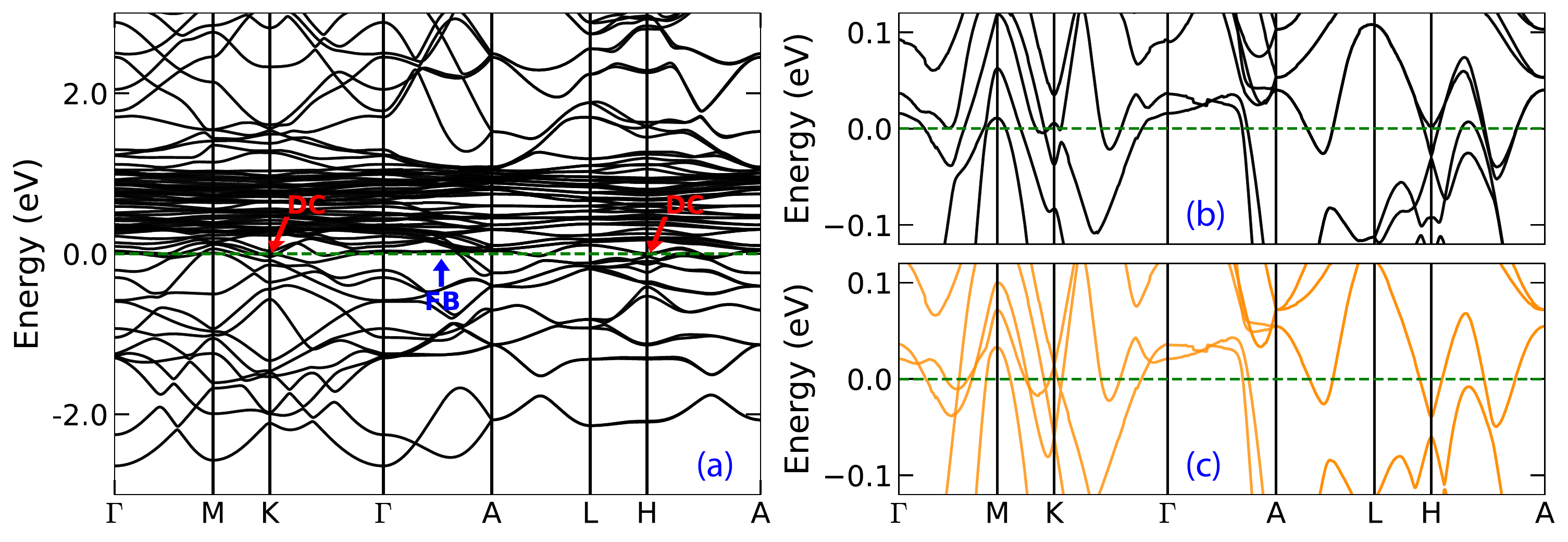}
\caption{(Color online). (a) and (b) Electronic band structures of $\alpha$-Ce$_{3}$Al obtained via DFT + SOC calculations. (c) Electronic band structure of $\alpha$-Ce$_{3}$Al obtained via DFT calculations. Here DC means Dirac cone and FB is an abbreviation of flat band. The horizontal dashed line denotes the Fermi level. Note that the Dirac cones or gaps at $K$ and $H$ points should be tuned by spin-orbit coupling. \label{fig:soc}}
\end{figure*}

\begin{figure*}[ht]
\centering
\includegraphics[width=\textwidth]{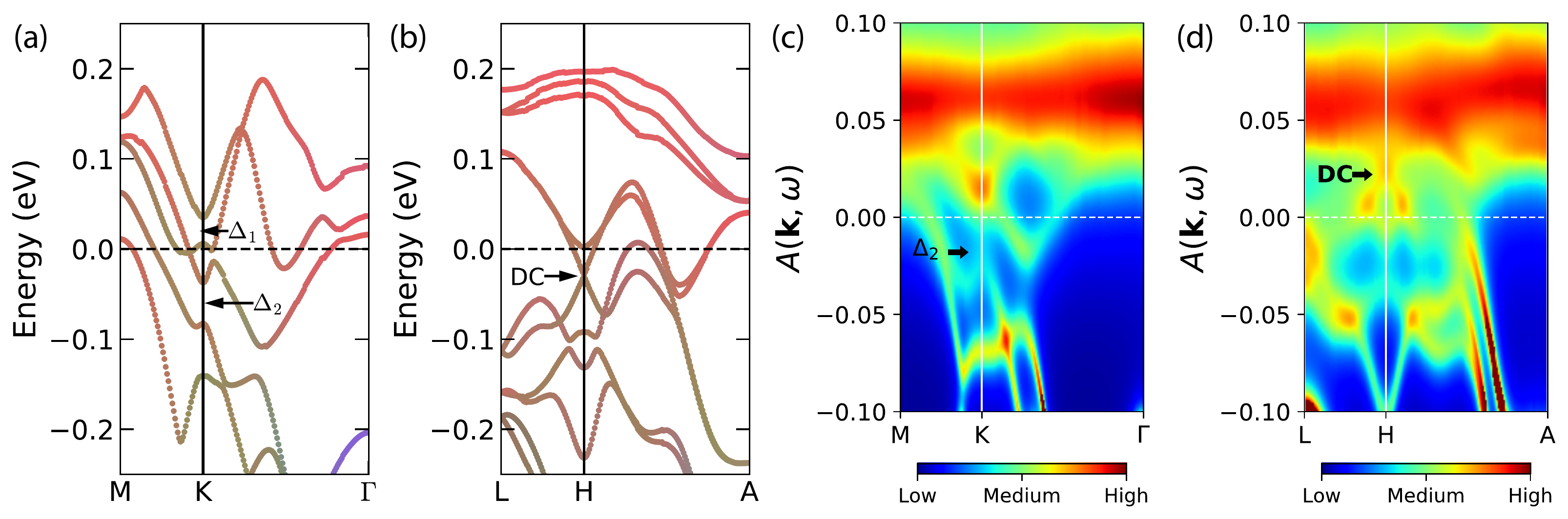}
\caption{(Color online). (a) and (b) DFT + SOC band structures for $\alpha$-Ce$_{3}$Al in some selected high-symmetry directions. Here, colors are used to distinguish different orbital characteristics (Red: Ce-$4f$; Green: Ce-$5d$; Blue: Al-$3p$), $\Delta_1 \sim \Delta_2$ mark the Dirac band gaps opened by SOC, DC means the Dirac cone. (c) and (d) Quasiparticle band structures for $\alpha$-Ce$_{3}$Al in the same high-symmetry directions as in (a) and (b). Here, $\Delta_2$ and DC mean the renormalized Dirac gap and Dirac cone, respectively. \label{fig:band}}
\end{figure*}

Very recently, the concept of kagome lattice and kagome metals have attracted quite a lot of attentions~\cite{PhysRevB.45.12377}. The kagome lattice or kagome pattern is a two-dimensional network of trihexagonal tiling. If conductive materials are made to resemble some kind of kagome lattice at the atomic scale, they are the so-called kagome metals. From the viewpoint of band topology, the signature of kagome metals is the coexistence of kagome-derived dispersionless flat bands and band crossings between two linearly dispersive bands (i.e. Dirac cones)~\cite{PhysRevLett.121.096401,Ye2019,Yin2018}. When the spin-orbit coupling is nontrivial, Dirac gaps are opened and massive Dirac fermions emerge~\cite{Ye2018}. Thus, the kagome metals provide ideal platforms to study the interplay between topology and electronic correlation. Up to now, most of the available kagome metals are correlated $d$-electron systems, containing Fe, Co, and Mn transition metal elements~\cite{Ye2018,PhysRevLett.121.096401,Ye2019,Wang_2020,Yin2018,Kim2018,Liu2018,Wang2018,Yin2019,PhysRevLett.124.077403,zl:2020,Kang2020,ws:2020,rc:2020}. To our knowledge, the $f$-electron kagome metals are still absent. It would be interesting to search kagome metals in rare earth and actinide compounds, especially in cerium-based heavy-fermion materials. 

Notice that the crystal structure of $\alpha$-Ce$_{3}$Al can be viewed as a sequence stacking of a kagome layer along the $c$ axis [see Fig.~\ref{fig:bz}(a)]. Its kagome layer consists of Ce atoms and the centers of the hexagons are populated by Al atoms. As compared to the other kagome metals~\cite{Ye2019,Ye2018,Zhangeaao6791,ws:2020}, there is no spacing layer in $\alpha$-Ce$_{3}$Al. Nevertheless, a question is naturally raised. Is $\alpha$-Ce$_{3}$Al a candidate of the heavy-fermion kagome metal? In other words, can we observe kagome-derived flat bands and Dirac cones simultaneously in the electronic structure of $\alpha$-Ce$_{3}$Al? In order to answer this question, we made a further analysis about the DFT + SOC and DFT + DMFT band structures of $\alpha$-Ce$_{3}$Al.

In order to exclude the effect of electronic correlation, we at first carried out DFT and DFT + SOC calculations for $\alpha$-Ce$_{3}$Al. Fig.~\ref{fig:soc} shows the calculated band structure. We see that, the extremely flat bands, which are largely attributed to the 4$f$ orbitals of Ce atoms, spread over the whole Brillouin zone. At the high-symmetry $K$ and $H$ points, there are some crossing points made by linearly dispersive bands. In Fig.~\ref{fig:soc}(a), some representative flat bands and Dirac cones (or Dirac gaps) near the Fermi level are annotated by arrows. Enlarged views for these subtle band structures are illustrated in Fig.~\ref{fig:soc}(b) and (c), and Fig.~\ref{fig:band}(a) and (b). The Dirac cones at $K$ point have a 4$f$ character. They are opened by the spin-orbit coupling to yield two Dirac gaps $\Delta_1$ and $\Delta_2$. Interestingly, the spin-orbit coupling leads to a converse consequence at the $H$ point. It closes a Dirac gap and recovers a Dirac cone [labelled by DC in Fig.~\ref{fig:band}(b)]. Anyway, all these characteristics suggest that $\alpha$-Ce$_{3}$Al is a candidate of the kagome metal with heavy fermions. These dispersionless flat bands and Dirac cones (or Dirac gaps) are indeed kagome-derived features, instead of a consequence of strong 4$f$ electronic correlation or $c-f$ hybridization effect.

Next, we wonder whether these kagome-related features will be modified by strong electronic correlation. In Fig.~\ref{fig:band}(c) and (d), the corresponding quasiparticle band structures obtained by DFT + DMFT calculations are shown. First of all, the kagome-derived flat bands are strongly renormalized. The band width is greatly reduced and the central energy level is shifted downward. Second, the Dirac gap $\Delta_1$ is destroyed. Though the Dirac gap $\Delta_2$ still survives, it is slightly shifted toward the Fermi level. Third, the Dirac cone at the $H$ point (DC) is pushed onto the Fermi level and touches the flat bands. The nearby band structure is quite incoherent. As a whole, the influence of 4$f$ electronic correlation on the kagome-derived bands can not be ignored in this case. Our results imply that there should be a competition between the kagome mechanism and the electronic correlation.   

\section{summary\label{sec:summary}}

In summary, we performed charge fully self-consistent DFT + DMFT calculations to study the detailed electronic structures of $\alpha$-Ce$_{3}$Al and $\beta$-Ce$_{3}$Al. The band structures, self-energy functions, and valence state histograms of them were determined. Both phases are typical heavy-fermion metals. Their 4$f$ electrons tend to be itinerant and strongly hybridize with the conduction electrons. The 4$f$ valence state fluctuations are comparable to those in $\alpha$-Ce. We estimated the single-impurity Kondo temperature $T_K$ and the coherent Kondo temperature $T^{*}_{K}$. We find that $T_{K} \gg T^{*}_{K}$, which agrees with the prediction of Nozi\'{e}res' exhaustion theory and suggests that both phases would retain the protracted Kondo screening state over a wide range of temperature. In addition, clear signatures of kagome metal, including dispersionless flat bands and linearly dispersive bands, are identified in the band structure of $\alpha$-Ce$_{3}$Al. Thus, it is concluded that the $\alpha$ phase of Ce$_{3}$Al is a candidate of heavy-fermion kagome metal. This material should provide a fertilizer playground to study the entanglement of topology, Kondo screening, and heavy-fermion behavior. Further experiments and theoretical studies to validate our predictions are highly desired. 

\begin{acknowledgments}
This work was supported by the Natural Science Foundation of China (No.~11874329, 11934020, and 11704347), and the Science Challenge Project of China (No.~TZ2016004).
\end{acknowledgments}


\bibliography{ceal}

\end{document}